\documentclass[preprint,proceedings]{rmaa}

\suppressfulladdresses
\usepackage{paralist}

\usepackage{psfrag,color}

\newcommand{\D}{\discretionary{}{}{}}
\newcommand{\rblr}{$r_\mathrm{BLR}$\/}
\newcommand{\kms}{km s$^{-1}$\/}
\newcommand{\hbbc}{H$\beta_\mathrm{BC}$\/}
\newcommand{\rfe}{$R_\mathrm{Fe \sc{II}}$\/}
\newcommand{\mbh}{$M_\mathrm{BH}$\/}

\SetYear{2007}
\SetConfTitle{The Nuclear Region, Host Galaxy and Environment of Active Galaxies}

\title{The Broad Line Region of Quasars}

\author{Paola Marziani\altaffilmark{1}, Jack W. Sulentic\altaffilmark{2}, and Deborah Dultzin\altaffilmark{3}
}
\altaffiltext{1}{INAF, Osservatorio Astronomico di Padova, Vicolo dell'Osservatorio 5,   I-35122 Padova, Italy; (paola.marziani\D{}@oapd.inaf.it).}
\altaffiltext{2}{Department of Physics and Astronomy, University
of Alabama, Tuscaloosa, AL 35487, USA; (giacomo\D{}@merlot.\D{}astr.ua.edu).}
\altaffiltext{3}{Instituto de Astronom\'\i{}a, Universidad Nacional Aut\'onoma de M\'exico,  Apartado Postal   70-264,  M\'exico D.F. 04510, M\'exico (deborah@astroscu.\D{}unam.\D{}mx).}

\shortauthor{Marziani, Sulentic \& Dultzin}
\shorttitle{BLR in Quasars}

\listofauthors{P. Marziani}
\indexauthor{Marziani, P.}

\abstract{The last decade saw  long-awaited improvements in our understanding of active galactic nuclei (AGN) spectral properties.  This contribution reviews some important observational results obtained from optical and UV data as well as  constraints on physical parameters that control the structure and dynamics  of  the Broad Line Region.}

\resumen{
A lo largo de la decada pasada se avanz\'o mucho en nuestra comprensi\'on de las
propiedades espectroscopicas de los NAGs (Nucleos Activos de Galaxias). Esta
contribuci\'on revisa algunos resultados observacionales importantes obtenidos
tanto en el \'optico como en el UV, asi como las restricciones que imponen sobre
la estructura y la din\'amica de las regi\'ones que emiten las l\'ineas anchas
 }

\addkeyword{Galaxies: Active}
\addkeyword{Quasars: General}
\addkeyword{Quasars: Emission Lines}
\addkeyword{Black Hole Physics}

\begin{document}
\maketitle

\section{Introduction}
\label{sec:intro}


``A  thousand spectra are worth more than one average spectrum" is
perhaps an appropriate modern  extension of the aphorism ``a
spectrum is worth a thousand pictures" -- at least for active
galactic nuclei \citep{cox2000,dultzinhacyanetal00}.  The ``one
thousand" spectra of a source are needed to obtain an estimate of
the spatial extent of the  broad line emitting region (BLR). The
time delay between line and continuum flux variations (i.e.
reverberation) is $\la$ 1 month, implying an angular size of just
0.1 marcsec at $z \approx 0.010$. Three orders of magnitude
improvement in resolution is still needed  to obtain a fuzzy view of
the BLR with direct, space-borne optical imaging.


Monitoring several lines emitted by ionic species of widely
different ionization levels  would advance another, yet small step
toward structure resolution. It is useful to separate lines into
high ionization (HILs: $\ga$ 40 $\div$ 50 eV): \ion{C}{4}$\lambda$1549,
\ion{O}{4}]+\ion{Si}{4} $\lambda$1400, \ion{He}{2}$\lambda$1640
\ion{He}{2}$\lambda$4686 and  low ionization (LILs: $\la$ 20 eV: Balmer lines,
\ion{Fe}{2}, \ion{C}{2}, \ion{Mg}{2}$\lambda$2800).
A major  effort in the 1990s established that HILs usually respond
with shorter time-scales than LILs  \citep[see e.g.,][ for NGC
5548]{koristaetal95}. Unfortunately, studies of response in narrow
radial velocity intervals across line profiles (i.e. 2D
reverberation mapping) have slowed since then \citep[but see
][]{kollatschny03}, not least because of  technical difficulties
demanding dedicated instrumentation \citep{horneetal04}.


\begin{figure*}[!t]  \vspace{0pt}
  \hspace*{-0.7cm}%
   \includegraphics[width=1.1\columnwidth]{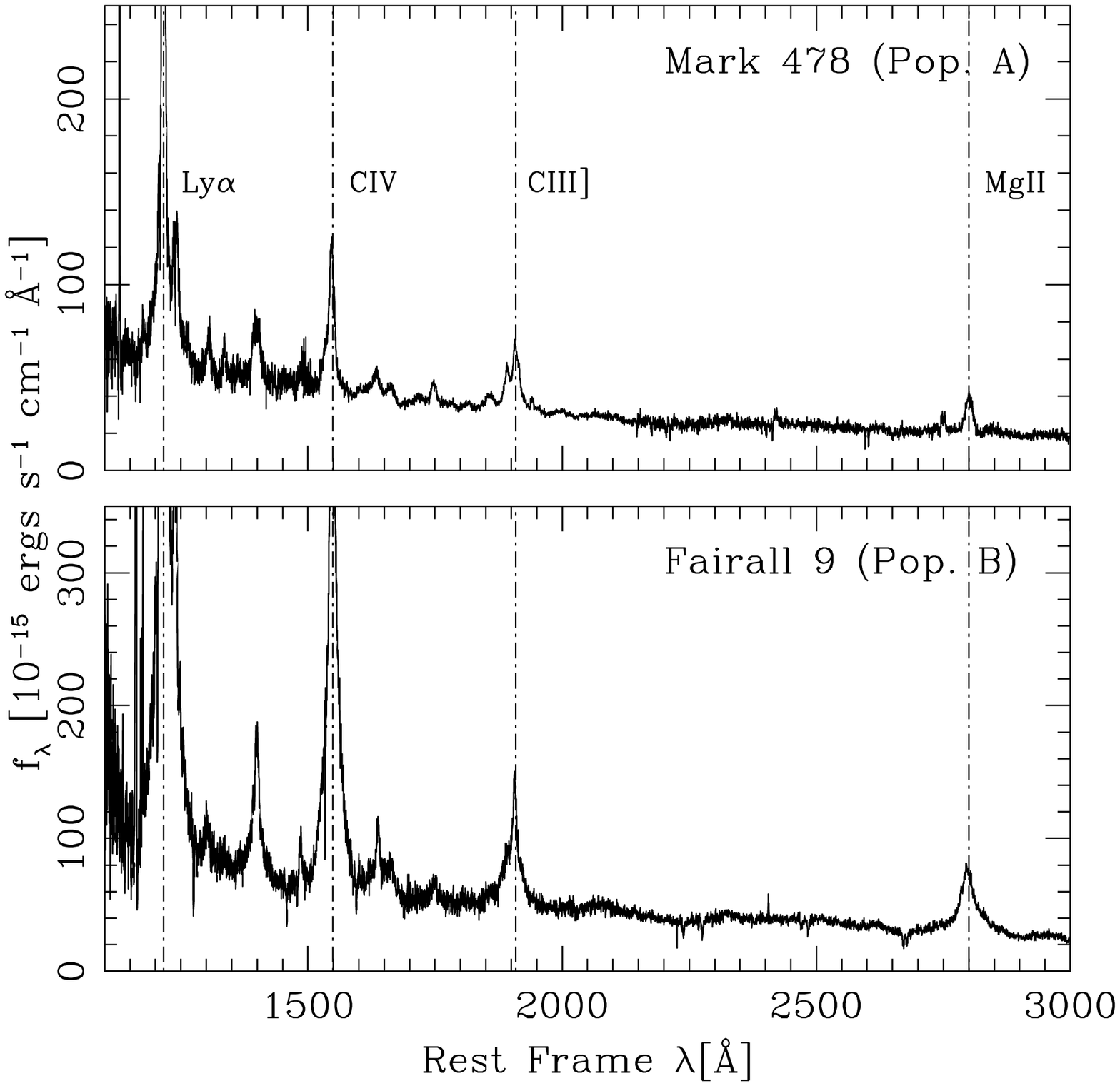},
   \hspace*{-0.7cm}%
  \includegraphics[width=1.1\columnwidth]{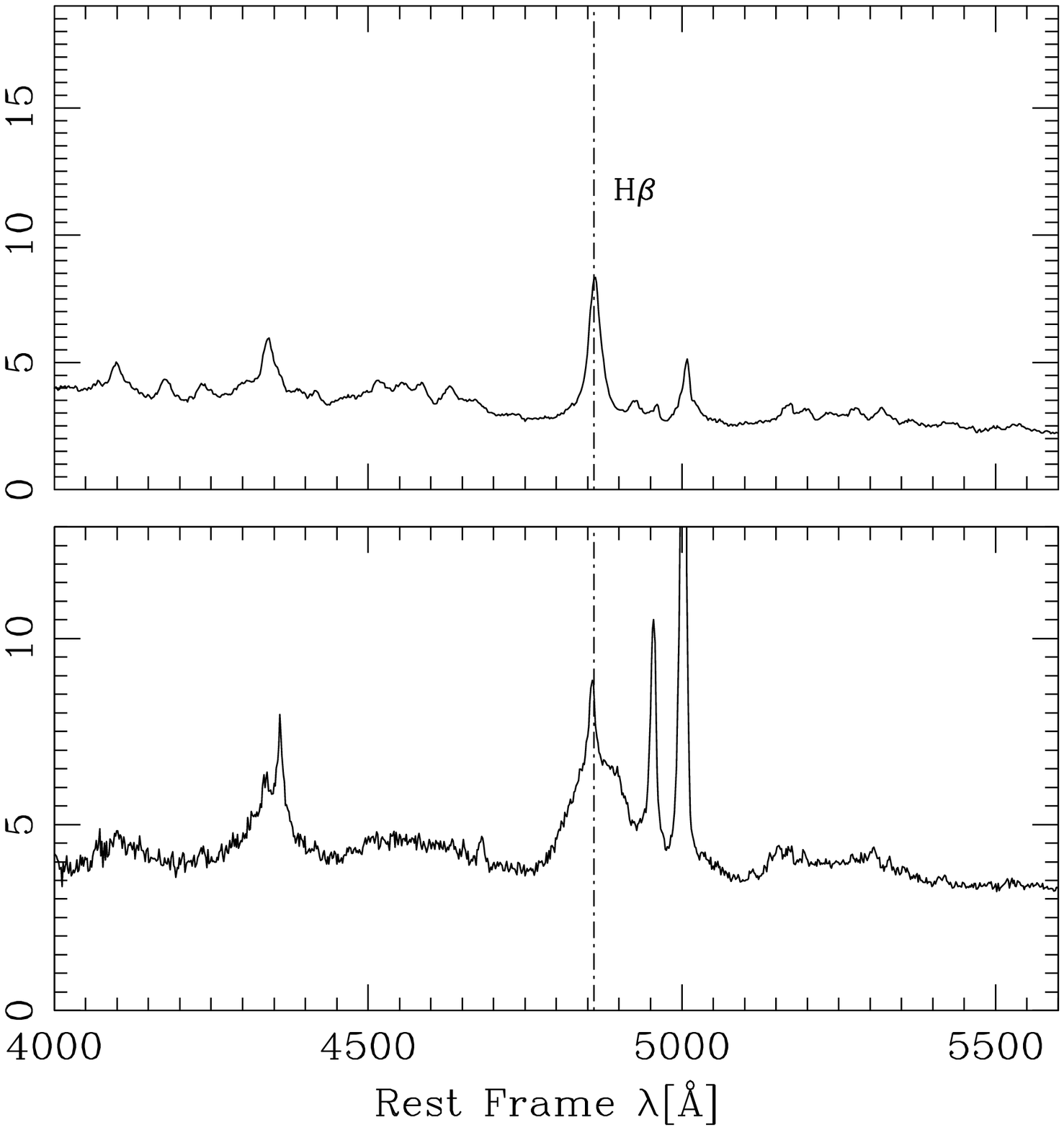}
  \caption{   HST/FOS (left panels) and optical, ground based spectra (right panels) of two representative sources: Mark 478 (Pop. A, top), and Fairall 9 (Pop. B, bottom). The strongest lines whose profiles are shown in the next figures are identified.
   }
  \label{fig:spec}
\end{figure*}

Emission line properties of quasars do not scatter randomly with
reasonable dispersion around an average, so that ``one thousand''
spectra are also needed to exploit  spectral diversity as well as
spectral variability. The range of  observed \hbbc\ FWHM spans more
than one order of magnitude, from $\approx 800$ \kms, to $\approx$
20000 \kms\ in some extreme sources.  The \ion{Fe}{2} prominence
parameter  \rfe\ (defined as the intensity of the \ion{Fe}{2} blend
centered at 4570 \AA\ normalized by the intensity of \hbbc) varies
from almost 0 to $\approx$1.5 -- 2.0 in large quasar samples
\citep{sulenticetal00}.


\section{Differences in BLR Structure and Kinematics}

Considering the diversity of quasar properties, it seems meaningful
to average spectra  in bins defined  within the so-called
``Eigenvector 1 optical plane'' of the 4D Eigenvector 1 parameter
space (see J.W. Sulentic' talk in these proceedings)  or to separate
at least narrower sources [FWHM(\hbbc) $\le$ 4000 \kms,
Population A] and B(roader) sources. There is evidence in favor of
two different BLR populations and of   a meaningful limit  at
FWHM(\hbbc) $\approx$ 4000  \kms\ \citep[see][]{sulenticetal07}. The
distinction between Population A and B may be more  fundamental than
either the RQ-RL or the NLSy1 vs. rest of quasars (BLSy1)
dichotomies. Several tests find no significant distributional
difference between NLSy1s and sources with FWHM(\hbbc) in the range
2000 -- 4000 \kms, suggesting that the 2000 \kms\ cutoff  is
artificial. Intriguing differences emerge if the Pop. A/B
distinction is applied. 

\begin{itemize}
\item  Pop. A  sources show Lorentzian \hbbc\ profiles, most often symmetric
or with slight blueward asymmetry, while Pop. B sources show double Gaussian
\hbbc\ profiles,  often redward asymmetric \citep[][see also 
Fig. \ref{fig:spec}]{sulenticetal02}.

\item The \ion{C}{4}$\lambda$1549  profile is most often  blueward
asymmetric and blueshifted in Pop. A, and it becomes a double
Gaussian, redward asymmetric or symmetric in Pop. B.  The
\ion{C}{4}$\lambda$1549 and \hbbc\ profiles appear to be more
similar in Pop. B. \citep[][ and references
therein]{sulenticetal07}.
\end{itemize}

Figure \ref{fig:spec} shows optical and UV spectra for
typical pop. A and B sources. A visual impression is that the ionization degree is higher for Pop. B \citep{marzianietal01}, with lower \rfe\ and more prominent HILs.  Kinematics and physical conditions may not
be fully unrelated, for example if both Pop. A and B sources  share a
low-ionization region responsible for most or all of the \ion{Fe}{2}
emission, while a  high-ionization, broader component is  dominant in  Pop. B only.

 \begin{figure*}[!t]  \vspace{0pt}
 \includegraphics[width=0.5\columnwidth]{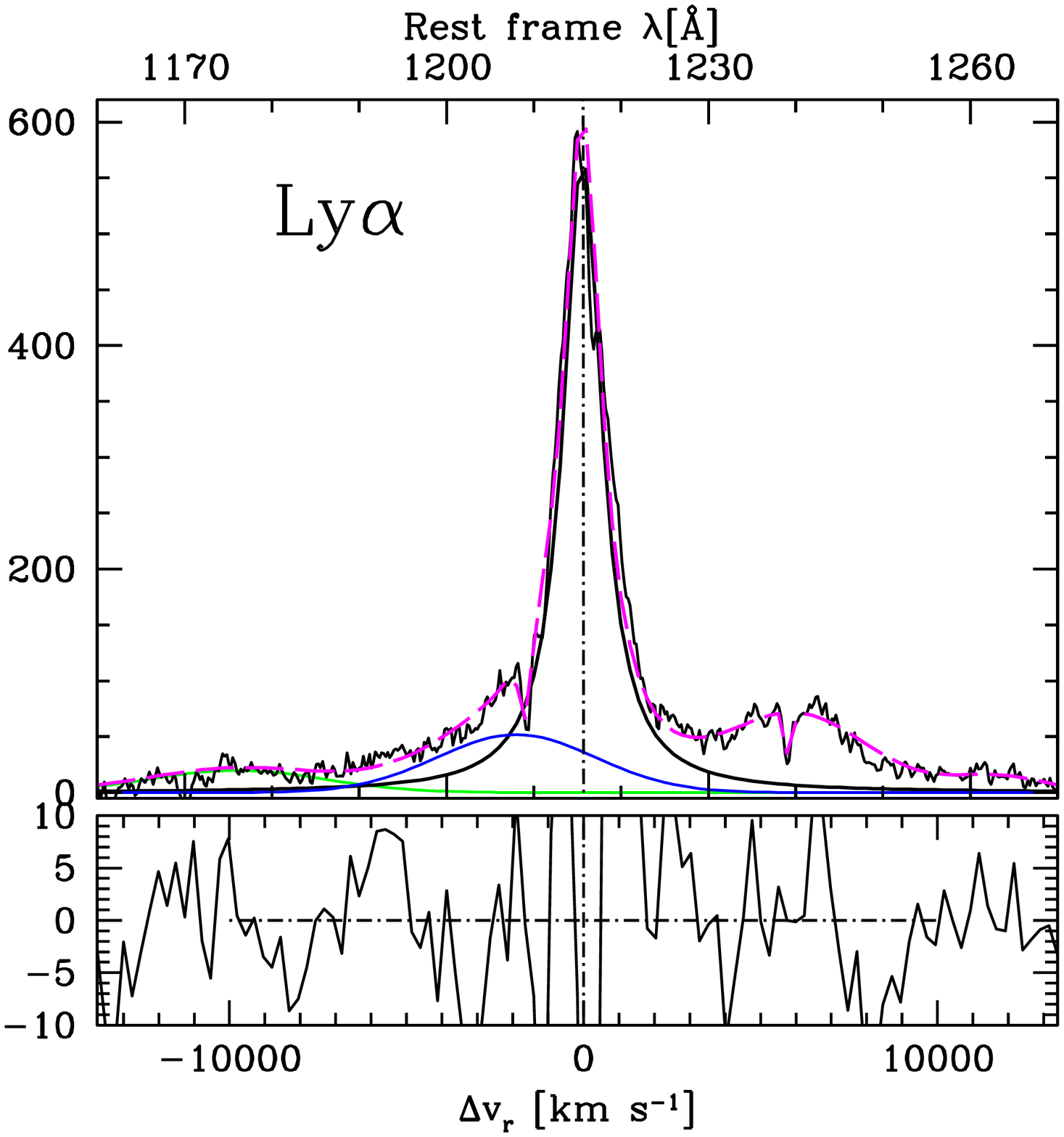},
\includegraphics[width=0.5\columnwidth]{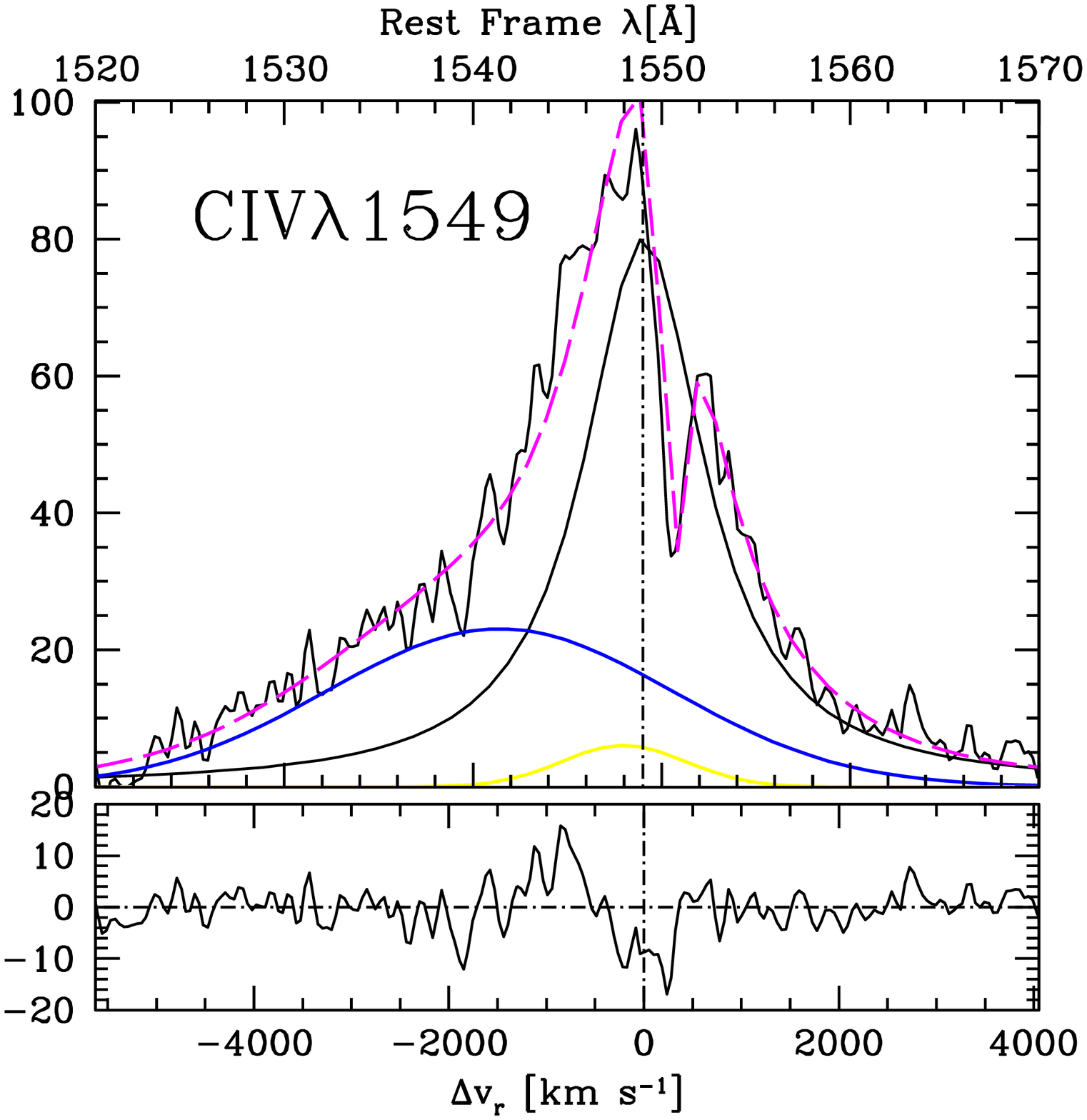}
\includegraphics[width=0.5\columnwidth]{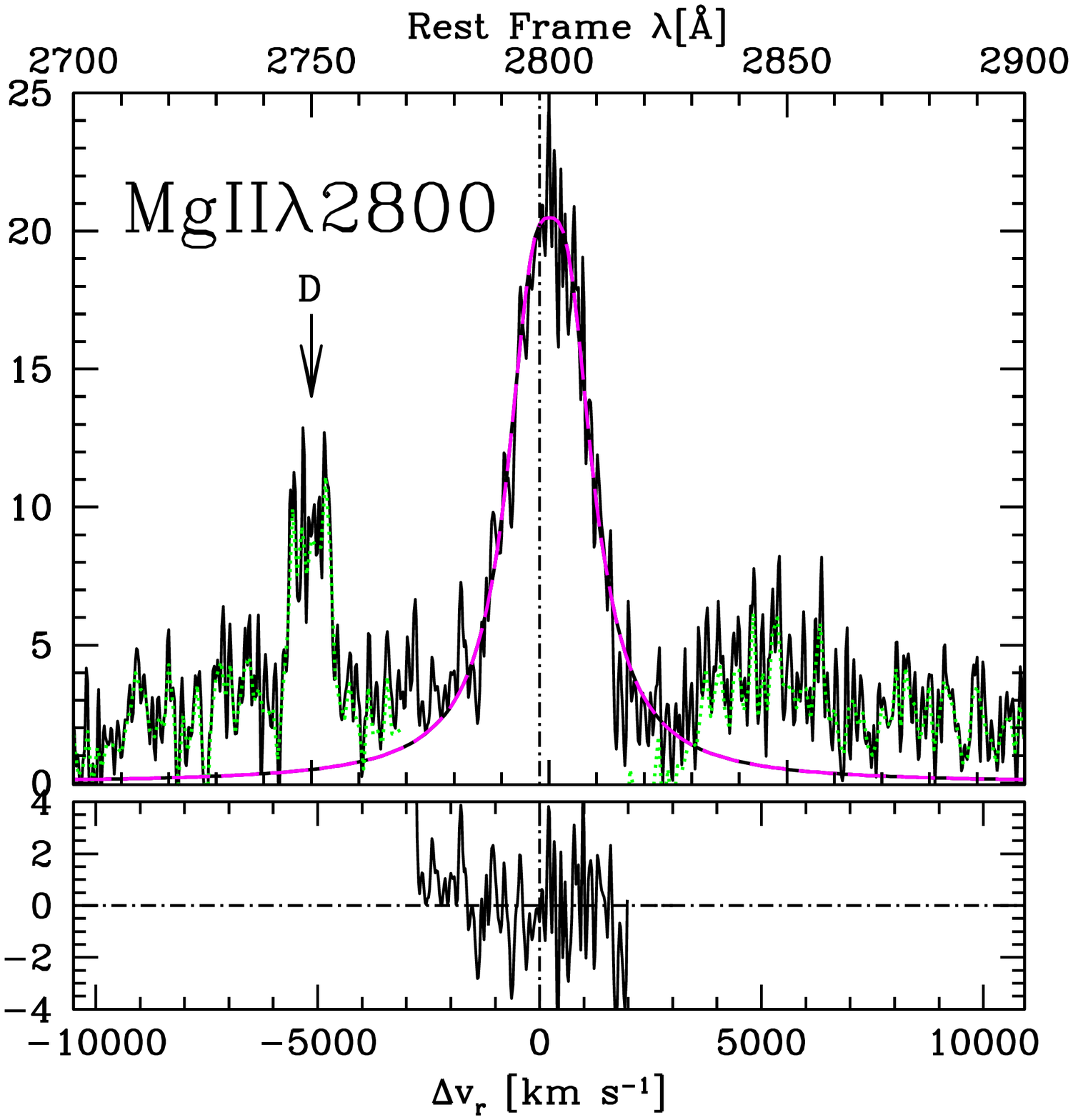}
\includegraphics[width=0.5\columnwidth]{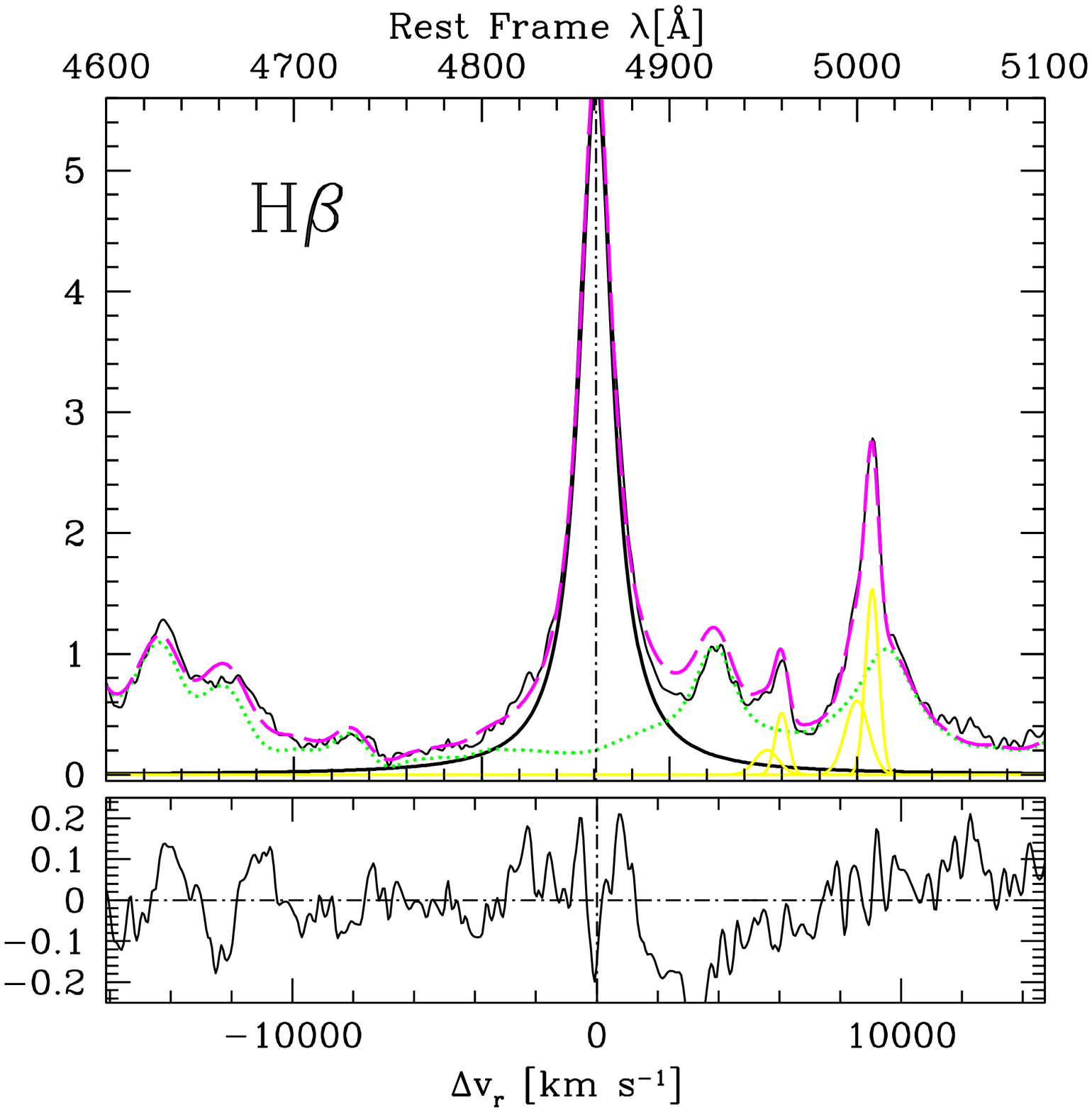}
  \caption{Profile analysis results for Mark 478, the prototypical Pop. A source considered in this study. Ordinate is rest-frame specific flux in units of 10$^{-15}$ ergs s$^{-1}$ cm$^{-2}$ \AA$^{-1}$, corrected because of Galactic extinction.  Dashed lines indicate sum of line model emission, dotted lines indicate \ion{Fe}{2} emission.  Due to the heavy \ion{Fe}{2} contamination we only fit the doublet
core of MgII.  See text for further details.
  }
\label{fig:m478}
\end{figure*}

 \begin{figure*}[!t]  \vspace{0pt}
\includegraphics[width=0.5\columnwidth]{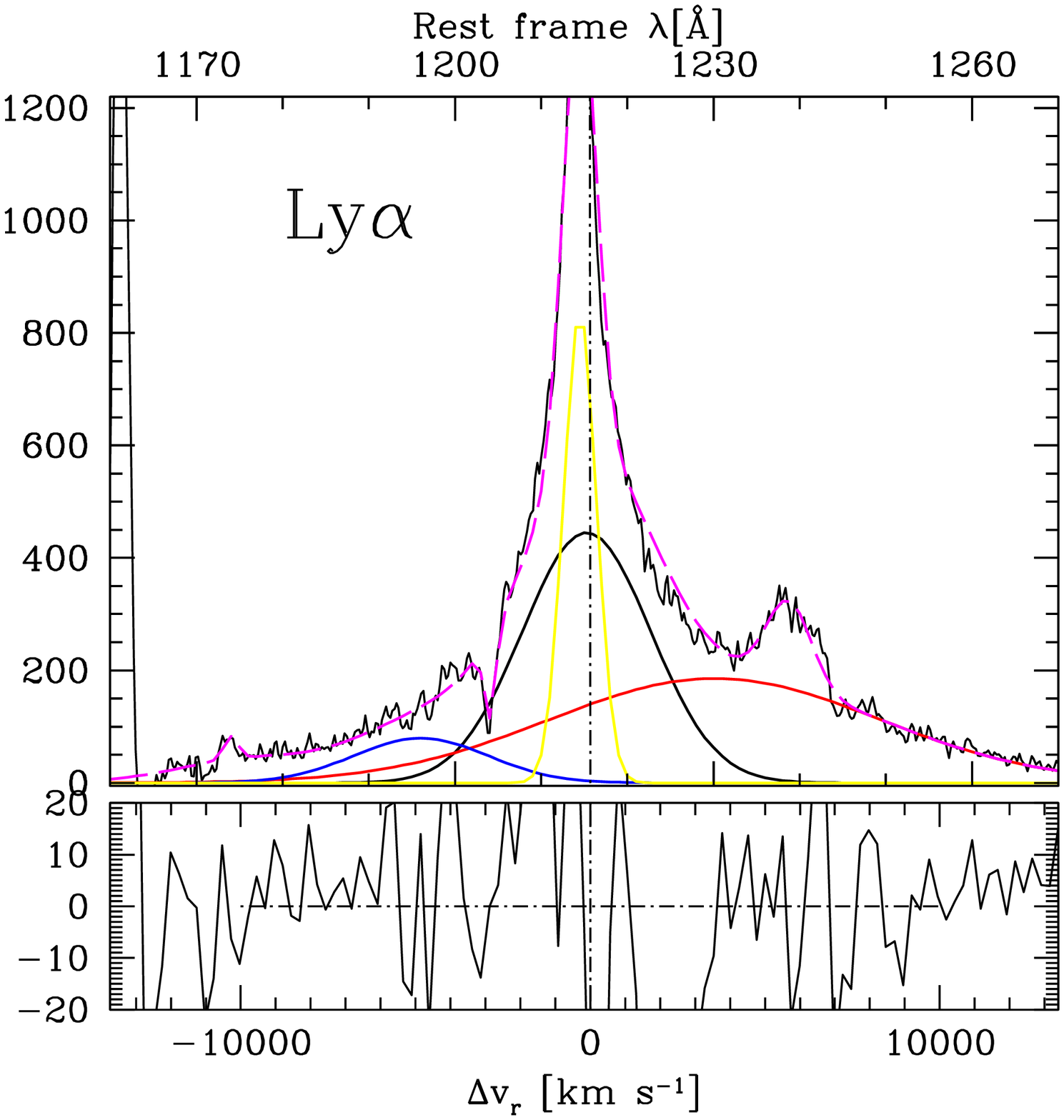},
\includegraphics[width=0.5\columnwidth]{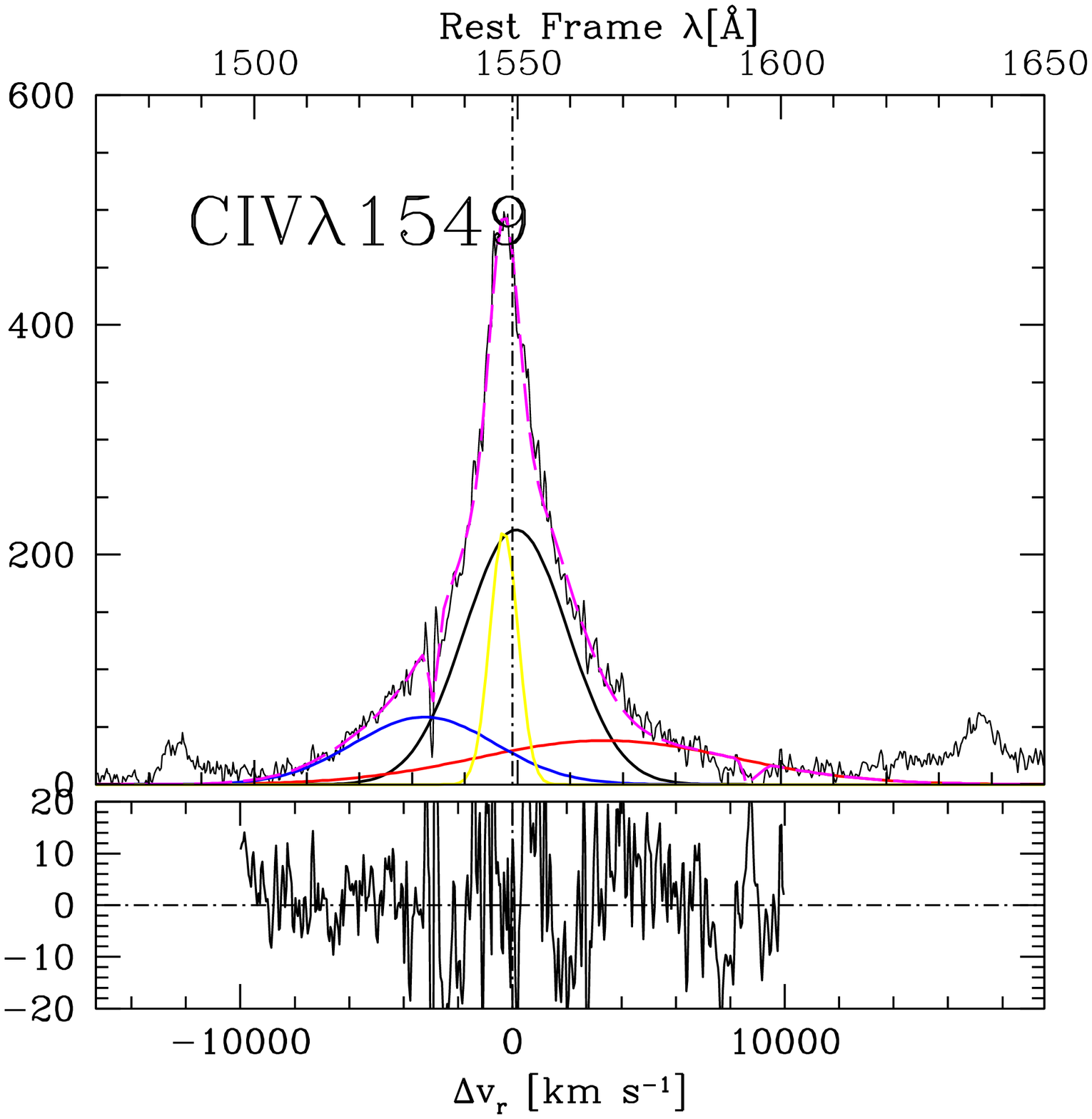}
\includegraphics[width=0.5\columnwidth]{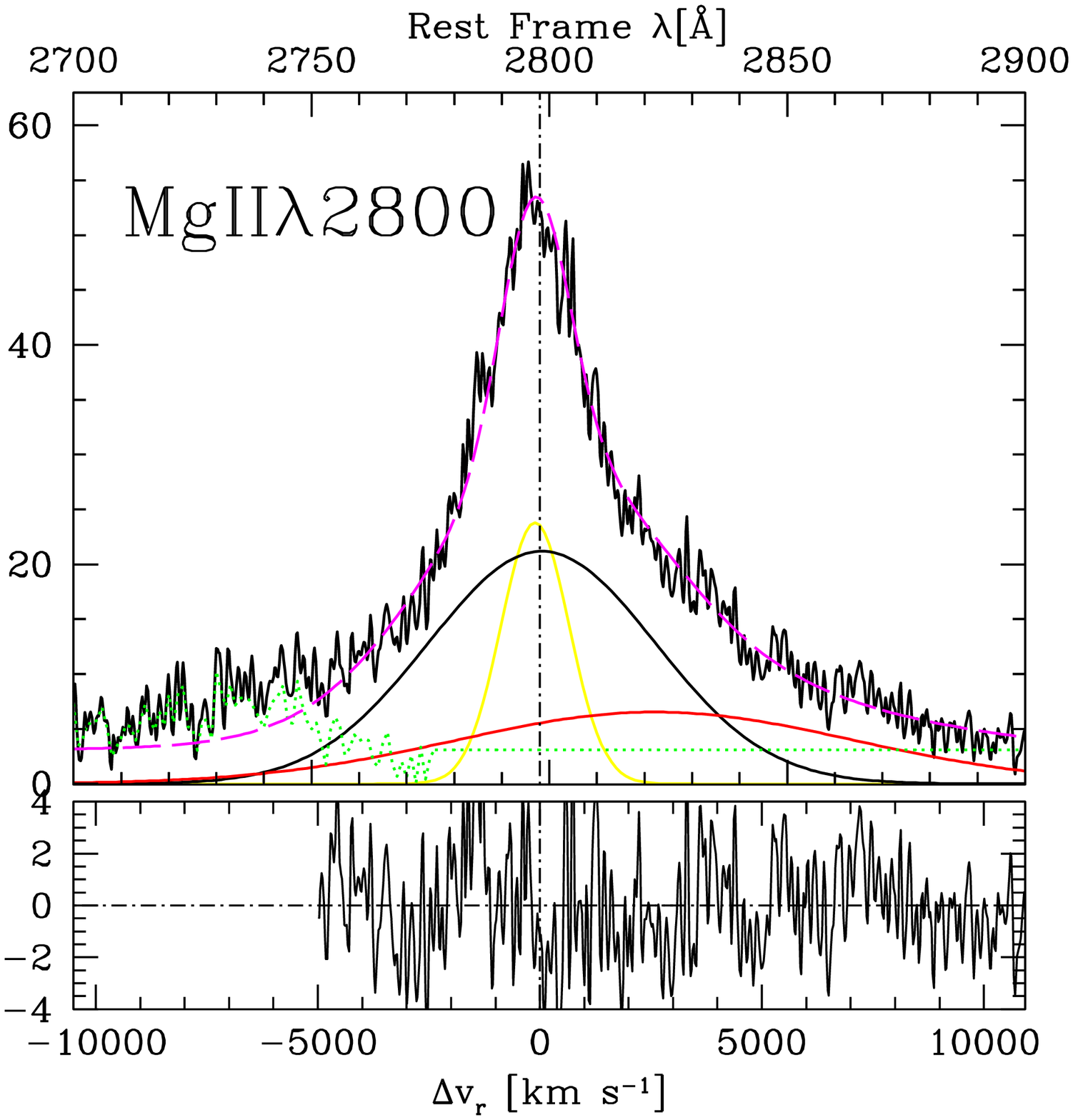}
\includegraphics[width=0.5\columnwidth]{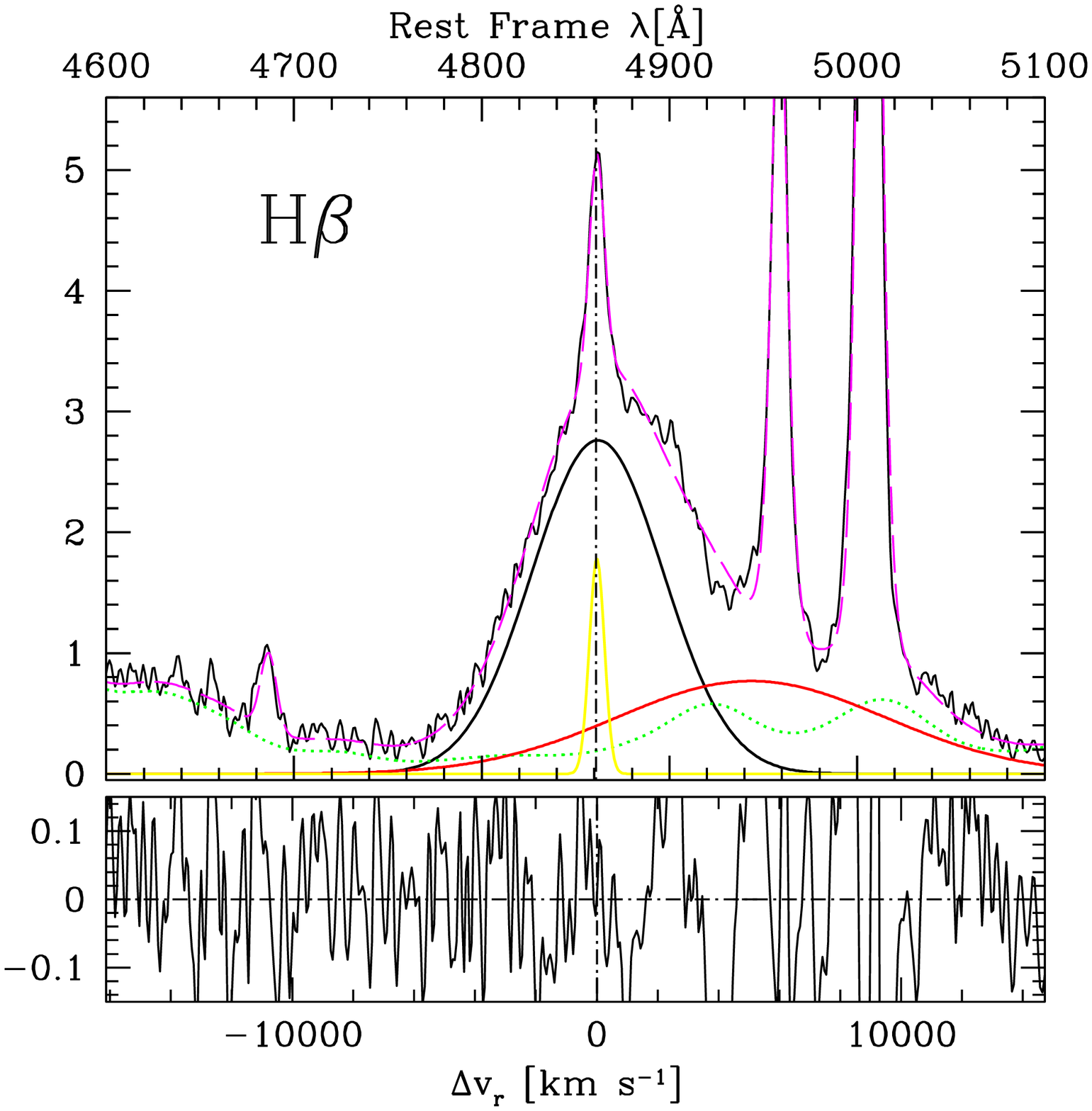}
  \caption{Profile analysis results of Fairall 9 (Pop. B prototype). Units and meaning of symbols are the same of the previous Figure. Due to the non-simultaneity of the observations, and to the rather variable behavior of  F~9 at the time the HST spectra were collected, it is not advisable to compare fluxes in the optical (H$\beta$) and UV  lines. }
  \label{fig:f9}
\end{figure*}

\section{UV/optical spectrophotometric comparison of 2 prototypical Sources }
\label{two}

We consider two sources in more detail, Mark 478 and Fairall 9,
which are representative of Pop. A and Pop. B respectively (Fig.
\ref{fig:spec}).  HST/optical data  allow us to analyze all the
strongest emission lines from Ly$\alpha$ to H$\beta$. 

\paragraph{Mark 478 (Fig.
\ref{fig:m478}) --} The \hbbc\ profile is well fit by a Lorentzian
function. If we assume that emission with the same profile  is also
present in \ion{C}{4}$\lambda$1549 then \ion{C}{4}$\lambda$1549 can
be decomposed into this component and a rather strong, blueshifted
component. A similar deconvolution very successfully reproduces the
Ly$\alpha$ profile.  No  blueshifted component is detected in
H$\beta$. \ion{Mg}{2}$\lambda$2800 is fit by a double Lorentzian.

\paragraph{Fairall 9 (Fig. \ref{fig:f9}) --} \hbbc\ and MgII$\lambda$2800 are  well fit with two Gaussian components, one narrower/unshifted and the other one much
broader and shifted to the red (``very broad component", VBC). Ly$\alpha$\ and \ion{C}{4}$\lambda$1549 may be equally well-fit by two components. However, the  VBC FWHM would be larger and the shift  lower than in the case of  H$\beta$\ and MgII$\lambda$2800, suggesting the presence of a third component. We prefer to fit Ly$\alpha$\ and \ion{C}{4}$\lambda$1549 as shown in Fig. 3:  a redshifted component analogous to the one identified in H$\beta$, and an additional blueshifted component.  \medskip

A tentative  nebular analysis can be carried out for the main
components identified in the profile decomposition using  {\tt CLOUDY} 
photoionization computations \citep{ferlandetal98}. {\tt CLOUDY 7.0} incorporates a 287-levels model of the \ion{Fe}{2} ion following \citet{verneretal99}. This is
especially useful since we are able to measure the \ion{Fe}{2}
emission for the 2000 -- 3000 \AA\ range (Fe$_\mathrm{UV}$), at least
normalized to Ly$\alpha$. We attempt to reproduce the observed line
ratios for the main components identified in the two sources always
assuming $N_\mathrm{H}  = 10^{23}$ cm$^{-2}$, standard AGN continuum,
and solar abundances. 

\paragraph{Mark 478: low ionization Lorentzian BC --} A very low ionization component accounts for most emission in the H$\beta$\ spectral region. Emission line ratios are best fit if ionization parameter is $\log \Gamma \la -2 $ and electron density $\log n_\mathrm{e} \ga 12$, where $n_\mathrm{e}$\ is expressed in cm$^{-3}$. High density and low ionization are especially needed to account for \rfe $\approx$ 1 and \ion{Fe}{2}$_\mathrm{UV}$/Ly$\alpha \approx 1$.  The very high $n_\mathrm{e}$\ is supported by the ratio \ion{C}{3}$\lambda$1176/Ly$\alpha \sim 0.1$.

\paragraph{Mark 478:  blueshifted  component --} The
blueshifted  component in Mark 478 appears to be less well constrained, partly because it is very weak in the H$\beta$\ range and  strong in the UV lines. Ionization is presumably high for the gas emitting this component ($\log \Gamma \sim -0.5 \div -1.0$) and density may be $\log n_\mathrm{e} \ga 10 \div 11$\ although  $N_\mathrm{H}$\ may be substantially lower than assumed in our explorative calculations.

\paragraph{F~ 9: low ionization Gaussian BC --} The  lower ionization component, with narrower profile, is responsible for {\em all} of the \ion{Fe}{2} emission, which is not negligible in F~9: Fe$_\mathrm{UV}$/Ly$\alpha \approx  $ 0.4, \rfe $\approx$ 0.7 for this component only (the 4DE1 \rfe\ includes all H$\beta$ flux [minus the narrow component] and is $\approx$ 0.4). The observed ratios are best explained if $\log \Gamma \sim -2$, and $\log n_\mathrm{e}  \sim 11 - 12$. These physical conditions are not much dissimilar from the ones deduced for the Lorentzian component of Mark 478 and, as expected,  the  \ion{C}{3}$\lambda$1176 line is detected.

\paragraph{F~9: high ionization, redshifted very broad Gaussian component (VBC) --} The absence of ``very broad" \ion{Fe}{2} emission, and the rather prominent  ``very broad" \ion{He}{2}$\lambda$ 1640 suggest  high ionization. This broader component can be accounted for by ionization parameter $\log \Gamma \sim -0.5 \div -1.0$, and moderate electron density $\log n_\mathrm{e} \sim 9.5 \div 10$. These parameters have been considered for a long time the canonical ones for the BLR.  

\paragraph{F~9:  blueshifted component --}  The high-ionization blueshifted component  used to fit the  Ly$\alpha$\ and \ion{C}{4}$\lambda$1549 profiles (Fig. \ref{fig:f9}) is interpreted as  analogous to the one observed in Mark 478. 
 \medskip 

Optical \ion{Fe}{2} emission is self-similar in almost all sources within common S/N  ($\sim 100$) and $\lambda/\Delta\lambda$\ ($\sim 10^3$) limits \citep{marzianietal03a}. The results obtained so far suggest that the similarity of the  optical \ion{Fe}{2} spectrum across Pop. A and B  is due to emission through photoionization of a very dense
medium. The connection between kinematics and ionization degree seems to be related to the relative prominence of the high and low ionization components in the HILs and LILs, since it is the low ionization component that produces all \ion{Fe}{2} emission.  How can we tentatively account for the different line profiles and
different average ionization conditions in Pop. A and B sources?

\section{Physical Parameters Behind the Observed Broad Line Diversity}


\subsection{Radio Loudness}

First, we consider whether a powerful, radio jet somehow affects the observational properties of the BLR.  To this aim, we compared median \hbbc\ profiles of 56 RQ  and 36 RL sources in the black hole mass  (\mbh) interval $8.5 < \log$Ê\mbh$< 9.5$ and
unrestricted bolometric luminosity-to-mass $L$/\mbh\ ratio \citep{marzianietal03b}. Most of these will
be Pop. B sources. The \hbbc\ profiles (after \ion{Fe}{2} and narrow
line subtraction) of RQ and RL sources are almost identical.  The
BLR (but not the Narrow Line Region!) does not seem to see whether a source is
radio-loud or radio-quiet.

\subsection{Luminosity}

VLT/ISAAC data provide IR spectra with resolution and S/N similar to optical data. We observed 50 sources to cover redshifted H$\beta$\ \citep{sulenticetal04,sulenticetal06}. Luminous quasars in the range 1.0$\la z \la 2.5$\ still show  Pop. A and Pop. B characteristics. LIL equivalent widths and \rfe\ do not appear to depend significantly on $L$,   even on a $\Delta m \sim$ 10 range. FWHM(\hbbc) is expected to increase $\propto 10^{-0.08 M_\mathrm{B}}$ if broadening of \hbbc\ is due to virialized gas
motions  with \rblr\ $\propto L^{0.7}$ \citep[see][for details]{sulenticetal04}.

\subsection{Gravitational Redshift}

The centroid at 0 intensity  $  c$(0/4) of \hbbc\  loosely correlates with \mbh;
the correlation is significant ($P \sim 10^{-3}$) because such a
large sample of $\sim$ 300 sources has been considered. If gas
motion remains virial down to the inner edge of the BLR there will
be a simple relationship between the  $  c$(0/4)  and
FWZI.  The amplitude of redward asymmetry depends on \mbh\ but
gravitational redshift seems to be statistically inadequate to
produce the $  c$(0/4) redshifts. The difficulty is even more
serious if the high-ionization VBC is considered
alone.

\subsection{Aspect Angle}

The angle $\theta$ between the line of sight and the radio jet axis  can
be  estimated equalling the observed X-ray flux at 1 KeV  to the
flux expected from the synchrotron self-Compton process acting on
radio photons  to obtain  the Doppler factor.  The  Lorentz factor
can be then retrieved from the apparent velocity if the source
is superluminal  \citep[following][and references therein]{sulenticetal03}.
Orientation matters, affecting  the FWHM(\hbbc) by a factor $\approx$ 2.
This technique can be applied only to quasars with detected superluminal
motion. There is yet no known way to retrieve  individual $\theta$s for
the rest of AGN \citep{collinetal06}.

\subsection{Eddington Ratio and Black Hole Mass}

The \hbbc\   profiles change strongly as a function of Eddington
ratio with profiles being generally Lorentzian if $\log L$/\mbh $\ga 3.9
$\ in solar units \citep{marzianietal03b} and Gaussian or
double-Gaussian below this limit.  The mass \mbh, estimated
according to the virial assumption, yields second order effects,
mainly in the line wings. The Eddington ratio also strongly affects
high ionization lines like  \ion{C}{4}$\lambda$1549: significant
 and large blueshifts ($\Delta v_\mathrm{r} \la -500$ \kms) are
confined to Pop. A sources only \citep{sulenticetal07}. These
results suggest that the Pop. A/B separation is physically motivated
and driven by Eddington ratio.

\subsection{Where is the Accretion Disk?}

Few ($\approx$ 2\%\  in the SDSS), very broad sources, with FWHM $\sim$ 6 times
the average FWHM of Balmer lines \citep{stratevaetal03}, show double-peaked
profiles that suggest accretion disk emission. To adjust the theoretical to the
observed profiles, non-axisymmetric   or warped disks are most often
required. A two component models for the LILs, a  disk (corresponding to
the high ionization component described in this paper) + a spherical one has
been successful for 12 AGNs, Pop. A and Pop. B. \citep{popovicetal04}, but
it is as yet unclear whether those results can be of general validity.
Eddington ratios of double-peaked sources are much lower than those of
typical type-1 AGN:  $\la$ 0.02 in 90\%\  \citep{wuliu04}. Old criticism based on line profile shape  for the wide majority of AGN is still standing \citep{sulenticetal90}: if only the accretion disk is emitting \hbbc, we expect that at the line base the centroid shift is due to gravitational and transverse redshift, but this is not always the case, not
even in Pop. B sources.

\section{Concluding Remarks}

Several BLR  physical and kinematical  properties seem to be
governed primarily by the Eddington ratio. It was not so clear ten
years ago. LILs properties are remarkably similar over a very wide
range of luminosity, even at very high $z$, and broad line profile
shapes (both LILs and HILs) show only second-order effects that can
be directly ascribed to \mbh. We have also learned with more
assurance that the BLR appears to be transparent to radio loudness.
This does not mean that RL and RQ samples show similar spectra: {\em
only} Pop. B RQ quasars and  lobe dominated RL quasar show very
similar spectra, with low \ion{Fe}{2} emission and broad \hbbc\ \citep{marzianietal96,dultzinhacyanetal00}. This
may point toward a parallel evolution for RL and part of the RQ
quasars: what makes them radio-different is an as yet unknown
parameter that  has little, if any, effect on their BLR.

We are still left with many conundrums concerning the actual origin
of the BLR gas, its dynamics and spatial disposition.
High-ionization gas (an accretion disk wind?), showing  non-radial,
outward motion and producing the blueshifted component observed in
Mark 478 and, to a lesser extent, in F 9, may decrease in importance crossing the boundary from Pop. A, where it is strong in the HILs, to Pop. B where it may become almost hidden by the redshifted, very broad component.  The central engine may sustain a most prominent outflow only if the Eddington ratio is relatively high, as
in the case of Pop. A sources.

Placing these considerations on a firmer observational basis
requires renewed efforts involving nebular diagnostics and line
profile analysis, as well as  the ability to understand how the
viewing angle affects observed  BLR parameters on a source-by-source
basis, if we cannot count on 2D reverberation mapping. 4DE1 offers
promise for decoupling source orientation from physics.


\begin{acknowledgements} PM wishes to thank the SOC for inviting her to speak in Huatulco. It was a valuable  opportunity to review past and current work of Deborah Dultzin  \&\ collaborators on the emitting regions in quasars, to which this paper -- given the vastness of the subject -- is limited.

\end{acknowledgements}


\end{document}